\documentclass[twocolumn]{aastex631}

\usepackage{newtxtext,newtxmath}   
\usepackage{wrapfig}

\begin{document}

\newcommand{\kms}{km~s$^{-1}$}
\newcommand{\ms}{m~s$^{-1}$}
\newcommand{\mss}{m~s$^{-2}$}
\newcommand{\dg}{$^{\circ}$}

\title{\large Meteoroid rotation and quasi-periodic brightness variation  of meteor light curves}

\author{\large Salvatore Mancuso$^1$, Dario Barghini$^1$ and Daniele Gardiol}

\affiliation{ Istituto Nazionale di Astrofisica, Osservatorio Astrofisico di Torino, via Osservatorio 20, Pino Torinese 10025, Italy \\ \email{salvatore.mancuso@inaf.it}}

\begin{abstract}
Meteor light curves are sometimes known to display flickering:  rapid, quasi-periodic variations in brightness. 
This effect is generally attributed to the rotational modulation of the ablation rate, which is caused by the time-varying  cross section area presented by a nonspherical rotating meteoroid to the oncoming airflow.
In this work we investigate the effects that the rotation of a meteoroid of given shape (spherical, cubic, or cylindrical) has on the meteor's light curve, given state-of-the-art experimental laboratory estimates of the drag and lift coefficients of hypersonic flow (Mach number $> 5$)  around various shaped objects.
The meteoroid's shape is important in determining these two forces, due to the different response of the drag and lift coefficients according to the angle of attack.
As a case study, the model was applied to a fireball observed on 2018 April 17 by the PRISMA network, a system of all-sky cameras that achieves a systematic monitoring of meteors and fireballs in the skies over the Italian territory.
The results show that this methodology is potentially able to yield a powerful diagnostic of the rotation rate of meteoroids prior to their encounter with the atmosphere, while also providing essential information on their pre-fall actual shapes.
\end{abstract}


\section{Introduction}

A meteoroid passing through the Earth's atmosphere, at a wide range of hypersonic velocities  from about 11 \kms\ to above 73 \kms\ (\citealt{Pena2024}), generates a visual fireball or bolide as a result of frictional heating induced by inelastic collisions with air molecules
(\citealt{Opik1958,McKinley1961,Ceplecha1998}). 
Part of the meteoroid kinetic energy is then transformed into visible radiation, and some of its momentum is also transferred to the air;   the resultant atmospheric drag decelerates the meteoroid 
(\citealt{Hill2004,Rogers2005,Vinkovic2007}). 

When the ablated material interacts through collisions with the atmospheric or evaporated meteoric atoms and molecules, it excites them, such that we observe the luminosity emitted in spectral bands due to the decay of these excited states (\citealt{Vondrak2008}).
Thermal ablation occurs intensively only after the meteoroid reaches the melting point, at the height of intensive vaporization (\citealt{Plane2012}) occurring in a range of altitudes between about 80 and 125 km. 
While most of the falling meteoroids are too small to survive thermal ablation, the surviving ones are also subject to fragmentation during their descent trough the atmosphere before reaching the ground
(\citealt{Ceplecha1998,Fisher2000,Rogers2005}). 
For the surviving meteorites, when ablation ceases, the body experiences a rapid exponential cooling with time, and the crust thus solidifies, impacting the ground at speeds of tens to  hundreds of \ms, according to their leftover mass. 
It is thus difficult to determine the actual physical characteristics of meteoroids entering the atmosphere since only a few of them are able to reach the ground, while small meteoroids ablate completely due to the removal of material from their surfaces.

The visual brightness of the falling meteoroid can be recorded by ground-based cameras.
Since the intensity of the light produced by a fireball is indicative of the almost instantaneous mass-loss rate, these light curves represent ideal indicators of the mode of meteor ablation and fragmentation, and even of the implied meteoroid structure.
Classically, the vast majority of light curve shapes can be represented by a steady rise to a maximum brightness followed by a steep drop.
However, faint meteors show a wide variety of  light curve shapes (\citealt{Fleming1993,Murray1999,Koten2001,Beech2003}). 
A few of them also display flickering,  a distinct quasi-periodic brightness modulation, whose amplitude typically changes with time and rarely exceeds 1 mag above or below the classical curve (\citealt{Beech2001}).
Flickering is present in about 5\% of observed fireballs and is a distinct phenomenon from the effect of flaring, which is due to stochastic meteoroid fragmentation.
\cite{Beech2000} identified the rotation of nonspherical meteoroids into the Earth’s atmosphere as the most likely underlying physical mechanism leading to flickering.

In this work we  examine the effects that the rotation of a nonfragmenting meteoroid of spherical and nonspherical shapes has on the meteor's light curve, given state-of-the-art laboratory estimates of the drag and lift coefficients of hypersonic flow (Mach number $> 5$)  around objects of various shapes  (spherical, cubic, and cylindrical).
We   show that the meteoroid's shape has an important impact on these two aerodynamic forces due to the different response of the drag and lift coefficients according to the body's angle of attack.
In particular, this analysis will yield an apt method to measure the rotation rate of meteoroids prior to their encounter with the atmosphere by examining the modulation of their brightness and speed dependence on height, while also providing essential information on the actual shape of meteoroids.

This paper is organized as follows. 
In Section 2 we present the PRISMA network, a system of all-sky cameras that achieves a systematic monitoring of meteors in the skies over the Italian territory.
Section 3 develops the time-dependent equations describing the meteoroid kinematics and ablation and discusses state-of-the-art laboratory estimates of the drag and lift coefficients of hypersonic flow around various shaped objects.
In Section 4, as a case study, we apply the model to a specific fireball observed on 2018 April 15 by the PRISMA network.
A discussion and our conclusions are  presented in Section 5.

\section{The PRISMA network}

The PRISMA network (\citealt{Gardiol2016,Gardiol2019}) is a system of tens of all-sky cameras that has been active since 2016; its  aim is to achieve a systematic monitoring of meteors and fireballs in the skies over the Italian territory.
PRISMA is a part of the international collaboration initiated by the Fireball Recovery and InterPlanetary Observation Network (FRIPON) project (\citealt{Colas2014,Jeanne2019,colas2020}). 
As a partner of the FRIPON collaboration, PRISMA shares the same technology of the network. 
Each station is equipped with a CCD camera (6 mm diagonal, $1296 \times 966$ pixel) coupled with a short focal lens objective (1.25 mm) in order to obtain an all-sky field of view.
The camera, connected to a Linux operating mini-PC via LAN and controlled by the open-source FREETURE software (\citealt{Audereau2014}), is operated at 30 frames per second (1/30 s exposure time) in order to sample the meteor trail at a suitable rate. 
The meteor detection is triggered locally in each node by a frame difference method, and cross-correlated with respect to the data of other nodes to check for multiple detections of the same event. 
The video stream of the detected meteor is saved locally as FITS files and, in the case of a multiple detection, the stream is collected by the FRIPON central server located at the Laboratoire d’Astrophysique de Marseille (LAM). 
Station monitoring, network security, software maintenance, real-time data processing, and sharing tasks are in charge of FRIPON and LAM teams. 
PRISMA data are synchronized and stored at the Italian Center for Astronomical Archives (IA2) INAF archiving facilities in Trieste (\citealt{Knapic2014}). 
The PRISMA reduction pipeline is implemented in IDL8 v8.7 and MATLAB9 Release 2015b and it was developed at the Osservatorio Astrofisico di Torino and Osservatorio di Astrofisica e Scienze dello Spazio in Bologna, which are part of the Italian National Institute for Astrophysics (INAF).
The astrometric calibration of the all-sky cameras of the PRISMA fireball network has been thoroughly described in \cite{Barghini2019}.
To date, PRISMA has observed hundreds of meteors, and even helped to locate the fragments of two meteoroids in Italy near Cavezzo, on 2020 January 4 (\citealt{Gardiol2021,Boaca2022,Bizzarri2023}), and near Matera, on 2023 February 14.

\section{Model}

The dynamics and ablation of a self-similar nonfragmenting meteoric body  of mass $m$ traveling through the Earth's atmosphere at a speed $v$ can be modeled by means of two kinematic differential equations that describe the change in  velocity (deceleration or drag) and mass loss (ablation) of a meteoroid traveling at hypersonic speed  (e.g., \citealt{Bronshten1983}):
\begin{equation}
\begin{aligned}
   & 
    \frac{dv}{dt} = - \frac{1}{2m} C_D S \rho_a v^2 +  g\sin{\theta}, \\
\end{aligned}
\end{equation}
\begin{equation}
\begin{aligned}
   &     \frac{dm}{dt} = - \frac{1}{2} \frac{C_H}{Q} S \rho_a v^3. \\
\end{aligned}
\end{equation}
In the above equations, $\rho_a$ is the density of the atmosphere through which the particle moves, $v$ is the magnitude of the velocity vector $\mathbf{v}$, $\theta$ is the inclination angle between $\mathbf{v}$ and the perpendicular to the gravitational acceleration $\mathbf{g}$, and $S$ is the middle cross sectional area of the frontal bow shock presented to the oncoming airflow and projected in the direction perpendicular to the flow.
In eq. (1) the coefficient of drag $C_D$ describes the contribution on falling meteoroids of drag forces $F_D = \frac{1}{2}C_D \rho_a v^2 S$ opposite to the motion.
This aerodynamic coefficients critically depends on the shape of the meteorite, but also on its speed and size and the properties of the atmospheric environment (and thus the flow regime) through the Reynolds ($Re$) and Mach ($Ma$) numbers (e.g., \citealt{Charters1945, Bailey1972, Henderson1976, Spearman1991}). 
A thorough discussion on flow regimes appropriate for meteor studies can be found in \cite{Moreno2018}.

In eq. (2) $C_H$ is a dimensionless heat transfer coefficient, which measures the efficiency of the collision process in converting kinetic energy into heat, while $Q$ is the heat of ablation  or the energy required to ablate a unit mass of the meteoroid, which is a function of the material type and the specific process of ablation.
According to \cite{Bronshten1983}, based on the best fit to the observational data, this value is about $C_H \approx 0.1$ or even lower by about an order of magnitude (e.g., \citealt{Gritsevich2011}), although this coefficient actually depends on speed $v$, altitude $h$, atmospheric density $\rho_a$, and body size.
On the other hand, a significant uncertainty is associated with this value (e.g., \citealt{Johnston2018}), and  in many works  $C_H$ is not even assumed, being identified as an integral part of a  mass-loss parameter $\beta$ (see \citealt{Gritsevich2007,Gritsevich2009}).
The value of  $Q$  is usually taken to be constant and equal to the heat of vaporization; a typical value for an ordinary chondrite meteoroid of density $\rho_m = 3.5\cdot 10^3$ kg/m$^3$ is $Q \approx 8\cdot 10^6$ J/kg (\citealt{Chiba1993}).
The ratio of the two parameters $C_H$ and $Q$ corresponds to the   ablation parameter, defined as $C_{ab} \equiv C_H/Q$ (e.g., \citealt{McMullan2019,Brykina2020,Bragin2021}).
We note that while $C_{ab}$ is clearly not fixed with velocity and radius, since this parameter is only poorly constrained in the literature, we   follow previous works and assume that it remains constant during the meteoroid's fall, at least in the range of heights where the fireball was observed.
In our model, $C_{ab}$ is  thus left as a free parameter.

The assumption of negligible lift for meteoroids is usually reasonable for first-order evaluations of aerodynamic effects on quasi-spherical falling bodies.
However, if the shape of the meteoroid is nonspherical or elongate, a lift force  $F_L = \frac{1}{2}C_L \rho_a v^2 S$ is expected when the object moves at an angle of attack $\alpha$ (calculated with respect to the semimajor axis of the elongated object) with respect to the resultant velocity vector $\mathbf{v}$ of the body.
The coefficients of lift $C_L$ also depends on the shape of the meteorite and on the Reynolds ($Re$) and Mach ($Ma$) numbers.
A maximum value for $C_L$ of order $\sim 0.001$ has been estimated by \cite{Passey1980} on the basis of the observed final distribution of impact craters, but this upper limit strictly applies only to massive ($> 100$ kg) meteoroids, while it is expected to be more pronounced for smaller meteors with shallow entry angles. 
For nonspherical bodies, the above system of differential equations  is thus necessarily completed by adding two other differential equations that describe the variations in the inclination angle $\theta$ with time along a meteoroid's trajectory and the variation in a meteoroid's height above the Earth's surface $h$ of radius $R_E$ as a function of time:
\begin{equation}
\begin{aligned}
   & \frac{d\theta}{dt} = \left(\frac{g}{v} - \frac{v}{h + R_E}\right)\cos{\theta} - \frac{1}{2m}C_L S \rho_a v,\\
\end{aligned}
\end{equation}
\begin{equation}
\begin{aligned}
   & \frac{dh}{dt} = - v \sin(\theta).
\end{aligned}
\end{equation}
While the gravitational force $\mathbf{F_G}=m\mathbf{g}$ is always directed towards the Earth's center, the aerodynamic drag $\mathbf{F_D}$ and lift $\mathbf{F_L}$ are respectively anti-parallel and perpendicular to the velocity vector.
The above system of four differential equations can be solved numerically by means of a fourth-order Runge–Kutta scheme if we provide an apt model for the atmospheric density $\rho_a$ as a function of altitude $h$ and appropriate boundary conditions.
Here we  assume an isothermal atmosphere $ \rho_a(t) = \rho_o\exp(-h(t)/H)$, where $\rho_o \approx 1.29$ kg m$^{-3}$ is the atmospheric density at sea level and $H \approx 7.16$ km is the constant scale height.
We note however that more reliable results can be  obtained by using real atmospheric measurements provided by national weather services for the times, heights, and locations of the specific fireballs under study (e.g., \citealt{Lyytinen2016}).

A different approach to the solution of the meteor's dynamics was proposed by \cite{Gritsevich2011}, based on the work by \cite{Stulov1995}.  
This method has the advantage that is entirely based on the interpretation of variations in  speed and height as consequences of braking and mass loss, with no assumption on meteoroid density or shape. 
In this approach the basic equations of meteor physics are rewritten using dimensionless parameters, and it is assumed that the  mass $m$ and cross section $S$ are connected through $S/S_e = (m/m_e)^\mu$, where $\mu$ is a coefficient ranging from 0 to 2/3, representing the effect of the body's change in shape, and the subscripts $e$ indicate the parameters at the entry to the atmosphere.
This is related to its rotation, which  prevents shape change and may distribute heat all around the
surface: a value of $\mu = 0$ corresponds to the case where the maximum heating and evaporation occurs in the front of the meteoroid, while $\mu = 2/3$ represents a uniform mass loss of the meteoroid over the entire surface.
In this different approach, the shape change coefficient, $\mu$, may be estimated through light curve analysis by fitting the meteor's luminosity.
In synthesis, this method allows us  to infer the initial mass, luminous efficiency, and ablation coefficient from a least-squares adjustment of the luminosity and velocity observations to the solution of the equations of atmospheric entry. 
Further details of the method can be found in \cite{Gritsevich2011}.

The amount of light produced by the meteor per second in the  instrument passband, or meteor luminosity $I$, is assumed to be proportional to the rate of change in the meteoroid's kinetic energy ${E_{kin}}$ and is described by the luminosity equation:
\begin{equation}
\begin{aligned}
    I = -\tau \frac{E_{kin}}{dt} = 
           -\tau \left(\frac{v^2}{2} \frac{dm}{dt} + m v \frac{dv}{dt}\right) 
    \hspace{0.5cm} [\rm J  \hspace{0.2cm} s^{-1}].
\end{aligned}
\end{equation}
This expression takes into account both the the mass-loss rate of ablation $\frac{dm}{dt}$ and the contribution from the deceleration $\frac{dv}{dt}$.
The coefficient $\tau$ in eq. (5) is the (dimensionless) instantaneous luminous efficiency, a parameter that yields the fraction of the meteoroid's kinetic energy that is converted into visible radiation.
The visual meteor luminosity, the portion of the total radiation in the optical bandpass (400--700 nm), can be finally expressed in terms of absolute magnitude $M_v$ given by
\begin{equation}
    M_v = -2.5(\log_{10}I - C)
,\end{equation}
where $C$ is a calibration constant, here taken as $C = 3.185$
(e.g., \citealt{Ceplecha2005,Gritsevich2011,Drolshagen2021}).

Since the typical speed in the atmosphere along the meteor's trajectory is much higher than 11 \kms, the flow regime is always expected to be hypersonic ($Ma = v/c_s > 5$) with a Reynolds number $Re$ above $\sim$$10^4$.
Depending on the speed of the object, its size, and mass-loss rate, a bow shock wave may  occur ahead of the meteoroid's trajectory.
The front bow shock separates a uniform flow from the field of perturbed flow in the shock layer around the point of braking or the stagnation point where the flow becomes subsonic.  
In the hypersonic case, the values of the drag and lift coefficients are relatively simple to estimate and do not critically depend on the atmospheric conditions.
In the literature the drag coefficient $C_D$ is usually set as a constant (between about 0.5 and 2), although it actually depends on many parameters, such as the shape of the meteoroid, its velocity, and the air density.
Since we model the meteoroid entering the atmosphere as a rotating body subject to both drag and lift effects, its shape is a major factor in determining these two forces, so that its cross sectional area $S$ has to be time dependent, thus somehow influencing the motion, also due to the different response of the drag and lift coefficients according to the angle of attack. 
In   recent decades, the drag coefficient has been calculated for several restricted ranges of $Re$ and $Ma$, and its estimation  has often relied on ad hoc interpolations between different regimes.

Spherical bodies are adequately studied aerodynamic test objects (\citealt{Charters1945,Hodges1957,Sighard1965,Bailey1972,Henderson1976,Spearman1991}).
These studies   experimentally ascertained that the drag coefficient $C_D$ of a spherical body does not depend on the Mach number $Ma$ when the flow is hypersonic and has an approximately constant value.
In general, for spheres traveling at hypersonic speeds in the continuum regime, the value of the drag coefficient is thus independent from $Re$ and $Ma$, and assumes an approximate value of $C_D \approx 0.92$ (\citealt{Sighard1965}).

\begin{figure}[t]
\centering
\includegraphics[width=8.6cm]{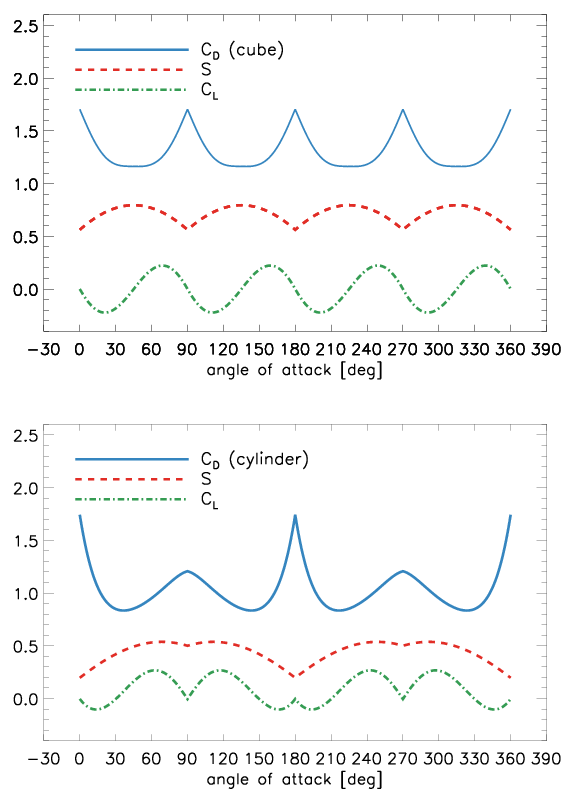}
\caption{Drag and lift coefficients as a function of the angle of attack for the cases of a rotating cube (upper panel) and a rotating cylinder (lower panel), as derived by \cite{Seltner2019,Seltner2021}. The red curve in the middle of each panel shows, in arbitrary units, the varying cross sectional area.}
\label{Fig1}
\end{figure}

The drag coefficient for nonspherical objects has also been a field of study for several decades. 
Theoretical models for the drag force on objects have however been derived only for a limited range of conditions, often subject to empirical limitations.

Fewer studies have been carried out on
 cube-shaped bodies (\citealt{Sighard1965,Hansche1952,Gerasimov2015}).
In particular, \cite{Hansche1952}   studied the dependence of the Mach number on the aerodynamic drag coefficient for rotating cubes free-flying in sub-, trans-, and supersonic flows, and a rough convergent value of $C_D \approx 1.66$ for Mach numbers $> 4$ has been given by \cite{Sighard1965}.
These earlier studies, however, dealt only with the drag, whereas the lift was neglected.
Furthermore, the aerodynamic coefficients were not related to the orientation angles of the cubes. 
In addition, these works provide  little information about the aerodynamic behavior of cubes in hypersonic flows.

In the first half of the last century studies on the aerodynamic behavior over a wide range of Mach numbers of cylindrical objects with both blunt and streamlined ends were conducted (e.g., \citealt{Vennard1940,Rouse1946,Gowen1952,Sighard1965}). 
These studies evidenced that the drag coefficient significantly depends on Mach number and is affected by the forebody shape in axial flows at transonic and supersonic speeds. 
However, these earlier studies   tended to focus on the drag of cylindrical bodies in axial flows and crossflows at supersonic speed, while the issue of inclined right circular cylinders in hypersonic flows was neglected. 
\cite{Seltner2021} addressed this gap by publishing the results of experimental and numerical investigations of pitched cylindrical bodies in hypersonic flow, ascertaining both the aerodynamic drag and lift force.

It has been recently proven by \cite{Seltner2019,Seltner2021} that the aerodynamic drag and lift coefficients of rotating cubes and cylindrical bodies in hypersonic flow have a significant dependence on the body orientation.
\cite{Seltner2019}   calculate the aerodynamic coefficients of a rotating cube using the free-flight technique in a hypersonic flowfield.
A rotating cube of side $l$ yields a pitch-angle-dependent projected frontal area given by $S(\theta) = l^2(|\sin(\theta)| + |\cos(\theta)|).$
The drag coefficient $C_D$ is thus not constant, but a periodic function of the pitch angle $\theta$ (see Figure~\ref{Fig1}).
The drag coefficient for a pitch-angle-dependent projected frontal area ranges between a minimum of about $C_D = 1.16$ and a maximum of $C_D = 1.71$ at a pitch angle equal to zero.
Moreover, a periodic lift force also arises for nonzero angles, with a lift coefficient $C_L$ continuously ranging between values of about +0.2 and -0.2 with the characteristics of negative sine functions.
The peak of the lift coefficient for a rotating cube is thus about 12\% of the maximum drag coefficient.
\cite{Seltner2021} also calculated the drag and lift coefficients for a rotating cylinder in hypersonic flow, obtaining a detailed dependence of $C_D$ and $C_L$ on the angle of attack, also shown in Figure~\ref{Fig1}.
In particular, the drag coefficient of cylinders of length $L$ base area $S = \pi D^2/4$, where $D$ is the cylinder's diameter and $L = 2D$ shows a more important influence of the inclination angle than in the case of cubes, with $C_D$ almost doubling between a pitch angle of 0\dg\ and 90\dg. 
In the following section we  apply the above-described model, complemented with the theoretical drag and lift coefficients calculated by \cite{Seltner2019} and \cite{Seltner2021}, to a specific fireball observed by the PRISMA network on 2018 April 17.

\section{A case study: The meteor observed on 2018 April 17}

\begin{figure}
\centering
\includegraphics[width=8.6cm]{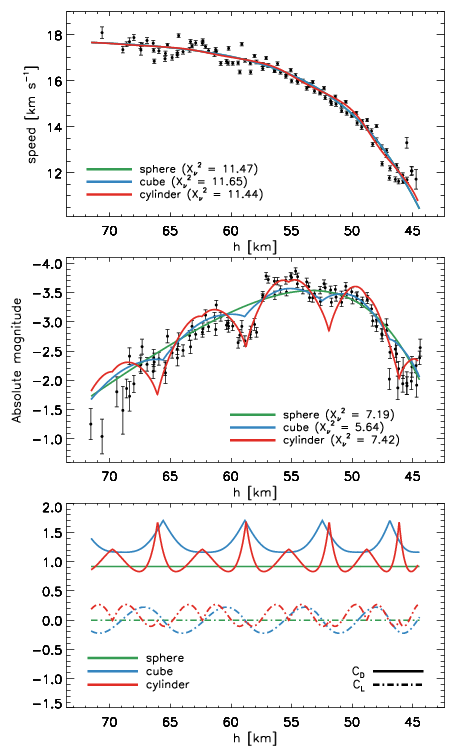}
\caption{Speed (top panel) and absolute magnitude (middle panel) vs height for the 2018 April 17 fireball, as seen from the PRISMA cameras.
Best-fit model curves for the three chosen meteoroid shapes (sphere, cube, and cylinder) are also shown superposed to the data. 
The prominent oscillations in the light curve are interpreted as being due to the asymmetric shape of the spinning meteoroid. 
The bottom panel shows the height dependence of the theoretical drag and lift coefficients for the three selected shapes.} 
\label{Fig3}
\end{figure}

On 2018 April 17, at around 19:45 UT, three stations of the PRISMA network (ITER04-Bedonia, ITER05-Piacenza, ITVA01-Lignan) detected a brilliant fireball in the skies of northern Italy.   
The meteor was tracked  from an initial height of 71.56 km to a final height of 44.45 km, with an initial velocity of about 17.720 \kms\ and a time of flight of about 1.97 s. 
The light curve (see middle panel of Figure~\ref{Fig3}), which reached a minimum absolute magnitude of -3.87, did not show sudden brightenings that might be attributed to important fragmentation episodes.

We solved the system of differential equations by using a standard fourth-order Runge-Kutta algorithm (RK4), as supplied in the Interactive Data Language (IDL), using a time step of 2 ms.
Runge-Kutta methods offer higher accuracy than the Euler integration method and are known for their accuracy, especially at higher orders. 
RK4, in particular, is widely used and allows a good balance between accuracy and computational efficiency.
We applied the model described in the previous section to this particular event as a case study to explore the implications of the changing aerodynamic coefficients of a spherical or nonspherical rotating body as a function of the angle of attack during its fall through the Earth's atmosphere.
For the ablating meteoroid, we considered rotating spherical, cubic, and cylindrical shapes, as discussed in the previous section, by using the state-of-the-art theoretical aerodynamic coefficients for hypersonic flows given by \cite{Seltner2019,Seltner2021}; we neglected the possible changes in shape due to ablation.
For modeling the absolute magnitude as a function of height, we had to choose a functional form for the luminous efficiency.

The coefficient $\tau$ is  not a well-known quantity, and it critically depends on other parameters, such as $m$, $v$, and even $\rho_m$.
The uncertainty on $\tau$ derives mostly from the fact that the majority of light emitted by fireballs is restricted to spectral lines that depend on the specific meteoroid's chemical composition (\citealt{Ceplecha1998,Campbell2012}).
On the other hand, \cite{Gritsevich2011} were able to compute $\tau$ with only very few assumptions, starting from the drag and mass-loss equations. 
In their approach, considering the change in the meteoroid’s speed and mass during its trajectory and taking the geometrical relation along the path of the meteor into account, they solved the formulas for the meteoroid’s dynamical behavior. 
The results were then compared to the observed meteor's drag rate and light curve, and $\tau$ was finally computed on the basis of this comparison.
Many values for $\tau$ are quoted in the literature according to the different assumptions the authors   made and the methods applied (e.g., \citealt{Verniani1965,Ceplecha1976,Halliday1996,Hill2005,Weryk2013}). 
All these studies found a dependency of $\tau$ on the speed $v$ of the meteoroid.
However, according to a recent study of luminous efficiency values determined with modern high-resolution instruments and obtained by directly comparing dynamic and photometric meteoroid masses, there is no clear dependence of $\tau$ on initial velocity (\citealt{Subasinghe2018}). 
Moreover, most nonfragmenting meteoroids appear to have luminous efficiencies of less than 1\% (\citealt{Subasinghe2018}).
Given the large uncertainty on the luminous efficiency $\tau$, this parameter is left as a free parameter in our model in order to match the observed absolute magnitude as a function of height.

We thus solved the system of differential equations presented in the previous section by using the theoretical drag and lift coefficients  given by \cite{Seltner2019,Seltner2021} that varied as a function of time in dependence on the supposed rotation of the meteoroid along the trajectory and the angle of attack.
The criteria for the  best fit  were a set of initial parameters (size of the body $h_0$, $C_{ab}$, and $\tau$) that best fit in a least-squares sense both the fireball's speed and the absolute magnitude dependence on height.
The initial size of the body $h_0$ was identified with the radius in the case of the sphere, the side length in the case of the cube, and the height in the case of the cylinder. 
For the cubic and cylindrical bodies two more parameters, the period of rotation $P_0$ and a phase factor $\phi_0$, were also introduced in order to model the modulation of the light curve of the fireball and of, at least in principle,   the speed during the meteoroid's fall.
The best-fit models for the three cases are displayed in Figure~\ref{Fig3}, together with an evaluation of the respective reduced chi-squared $\chi_\nu^2$.

Our best-fit model for the ideal spherical case, obtained by assuming a typical value for an ordinary chondrite meteoroid density ($\rho_m = 3.5\cdot 10^3$ kg/m$^3$), implies an initial instantaneous luminous efficiency $\tau = 0.134$\%, an initial radius $h_{0} = 0.0055$ m, and a constant ablation parameter $C_{ab} = 1.74\cdot 10^{-8}$ kg J$^{-1}$. 
Given the assumed meteoroid density, the initial mass was $m = 0.00247$ kg.
As for the two nonspherical cases, we found $\tau = 0.041$\%, $h_{0} = 0.0132$ m, $C_{ab} = 2.45\cdot 10^{-8}$ kg J$^{-1}$ for the cubic case, and $\tau = 0.077$\%, $h_{0} = 0.0184$ m, $C_{ab} = 1.89\cdot 10^{-8}$ kg J$^{-1}$ for the cylindrical case.
We also inferred a possible periodicity of $P_{0} = 1.778$ s for the rotating cubic body and a possible periodicity of $P_{0} = 0.964$ s for the rotating cylindrical body.
Interestingly, the best-fitting model for the cylindrical case clearly displays a peculiar slight modulation of the speed curve, which is, at least in principle, in view of more refined future fireball observations, potentially discernible from the observed speed dependence with height.
For the absolute magnitudes, by visual inspection, the curve corresponding to the rotating cylindrical shape appears to nicely follow the observed modulation although, perhaps counterintuitively, the minimum reduced chi-squared $\chi_\nu^2$ is actually achieved in the case of the spinning cubic shape, thus hinting to the possibility that an intermediate brick-shaped body rotating with a period of $P_{0} \approx 1.8$ s could have been responsible for the observed modulation.
In the bottom panel of Figure~\ref{Fig3}, for a better visual evaluation of the effects of the varying aerodynamic coefficients on the modulation of the speed and absolute magnitude, we also show the height dependence of the theoretical drag and lift coefficients for the three selected cases.

From the above case study it is thus clear that our model is able to satisfactorily explain the observed modulations in the light curves (and, at least potentially, in the speed curves) as being attributable to the nonsphericity of rotating falling meteoroids, even pointing to the possibility of determining,  with some degree of confidence, their probable shapes and rotation periods.

\section{Discussion and conclusions}

The drag and lift coefficients of nonspherical bodies are of crucial importance for the  correct computations of meteor-body parameters from observations.
In this work the effects on the modulation of an observed meteor's light curve due to the rotation of a rotating meteoroid of given shape that enters the Earth's atmosphere was investigated by applying state-of-the-art experimental laboratory estimates of the drag and lift coefficients of hypersonic flow (Mach number $> 5$)  around various shaped objects (spherical, cubic, and cylindrical), as given by \cite{Seltner2019,Seltner2021}.
The underlying major assumption is that the shape of the meteoroid does not change, and that during its descent through the atmosphere there are no episodes of fragmentation.
We   showed that the meteoroid's shape has an important impact on the drag and lift forces due to the different response of the respective time-dependent aerodynamic coefficients according to the given angle of attack.
These time-dependent coefficients are quite different from those generally used in the literature related to the studies of meteor events to calculate the motion of meteor bodies, which usually use an assumed guess value and/or a spherical shape.

Our model was applied to specific light curve data associated with a fireball observed on 2018 April 17 by the PRISMA network, which showed a clear modulation that, excluding a priori the possibility of fragmenting episodes, could be attributed to the rotation of a meteoroid.
Among the three cases, the cylindrical shape showed a peculiar speed modulation with height  and appeared to match, at least visually, the observed absolute magnitude dependence with height of the meteor. 
The cubic case, however,  was  the best-fitting of the three shapes, and displayed the best result for the reduced chi-square, hinting to the possibility that an intermediate brick-shaped body rotating with a period of $P \approx 1.8$ s might have been responsible for the observed modulation with height of the absolute magnitude.

As a final remark, we would like to point out that our enquiry, although it used accurate aerodynamic coefficients, was limited to specific perfectly shaped bodies, which may not represent the actual shapes of the vast majority of nonspherical meteoric bodies falling into   Earth's atmosphere.
For example, the rough surface of a quasi-spherical meteoroid would make   the value of the drag coefficient depart from its theoretical value of 0.92.
Moreover, even when not accounting for the body's surface roughness, if the meteor body is of a similar shape to those  modeled in this work (cubic or cylindrical),  the sharp edges of this nonspherical meteoroid  moving through the atmosphere are  expected to become rounded because of the effect of both melting and evaporation. 
This implies that the drag coefficient could considerably decrease with time due to the increasing rounding of the edges (e.g., \citealt{Zhdan2007}). 
Even with the above caveats, we demonstrated that our model yields a powerful means to estimate the rotation rate of meteoroids prior to their encounter with the atmosphere by examining the observed light and speed curves, while also providing essential information on their actual shapes.

\section*{Acknowledgments}
We thank the referee for carefully reading our paper and providing very helpful and thoughtful comments on our manuscript.
PRISMA is the Italian Network for Systematic surveillance of Meteors and Atmosphere. 
It is a collaboration initiated and coordinated by the Italian National Institute for Astrophysics (INAF) that counts members among research institutes, associations and schools. 
The complete list of PRISMA members is available at {\tt http://www.prisma.inaf.it}. 
PRISMA was partially funded by Research and Education grants from Fondazione CRT (years 2016, 2019, 2020, 2022, 2023). 
PRISMA data are hosted by the INAF research e-infrastructure project IA2 (Italian Center for Astronomical Archives).
The initial FRIPON hardware and software has been developed by the FRIPON-France core team under a French ANR grant (2014-2018).

\bibliographystyle{aa.bst}
\bibliography{biblio.bib}

\end{document}